# 光学误差对折射光线影响的理论分析与应用

黄卫东

中国科学技术大学地球化学与环境科学系

安徽合肥金寨路 96 号科大地学院，230026， email: huangwd@ustc.edu.cn

摘要：本文根据几何光学理论，提出了根据表面光学误差计算折射光线方向误差的通用表达式。结果表明，在离轴光学系统中，光学误差传递到折射光线上，存在误差放大效应。我们应用该表达式分析了 8 种太阳能聚光系统，给出了聚焦光线误差的计算表达式和计算结果。对点聚焦菲涅耳透镜，点聚焦抛物面玻璃反射镜， 使用玻璃反射镜的抛物槽，线聚焦菲涅耳玻璃反射镜，误差传递系数随反射点距离轴的距离的增加而增加，而棱镜全反射菲涅耳聚光镜则相反，随距离增加而减小。使用玻璃反射镜的定日镜主要与入射角和方位角相关。对槽式反射镜和定日镜，还存在 y 方向光学误差传递到光线 x 方向现象，表面坡度误差被放大到聚焦光线上。本文结论对改进聚光太阳能系统和其他光学系统性能具有重要指导意义。

关键词：光学误差，传递，聚光太阳能，折射光线

0、 引言

光学系统成像质量主要决定于光学设计、所使用的光学材料以及光学元件的加工和安装质量，光学元件表面加工质量是影响系统性能的关键因素之一。当一束光线射入有形状误差的光学表面，有入射位置不同于预计的问题，也有表面坡度与理论值不同的问题(Winston, Miñano et al. 2005)。从光学角度来看，通常的观点认为位置误差不重要，只需要考虑坡度误差(Rabl 1985)。光学元件表面误差的主要方面是表面倾斜角度与理想面的差别，一般将这种倾斜角度误差分解成相互垂直的两个方向上的误差，分别用$dw_x$和$dw_y$来表示，其方向的选取，与表面加工工艺相关产生的偏差分布有关，可以通过光学测量获得。人们发展了多种方法来测量这种误差，以评价和改进光学元件(Wendelin and Grossman 1994; Shortis and Johnston 1996; Ulmer, Heinz et al. 2009)。

根据几何光学原理，入射光经理想表面反射或折射后，反射或折射光方向遵从费尔马定理。实际光学表面存在误差，使反射或折射光偏离了理想方向，产生了误差。由于光学表面倾斜方向误差是随机的，反射或折射光线的分布也是随机，通常使用分布函数来描述。一般地，光学表面误差，及反射或折射光线的误差分布可用标准椭圆高斯分布表示(Butler and Pettit 1977; Bendt and Rabl 1981)：

$$B_{eff,Gauss}(\theta) = \frac{1}{2\pi\sigma_x\sigma_y}\exp(-\frac{\theta_x^2}{2\sigma_x^2} - \frac{\theta_y^2}{2\sigma_y^2})  \qquad (1)$$

我们对反射镜面误差到反射光线误差的传递进行了研究，结果表明，表面误差传递到反射光线上，存在放大效应(Huang 2011)。通常测量光学元件误差时，使用特定方向入射光线进行测量，得到的误差是特定入射光线下的反射或折射光线误差，不能代表一般情况下的反射或折射光线误差。例如，抛物面线聚焦槽式系统反射面上不同边缘角反射光线误差不仅与镜面横向误差和纵向误差相关，还与边缘角和入射角相关，其计算表达式为(Bendt, Rabi et al. 1979)：

$$\sigma_x^2 = 4\delta_x^2 + 4(\tan\lambda\sin\frac{\varphi}{2})^2\delta_y^2 \qquad (2)$$

当入射角接近 pi/2 时，横向反射光线的方向误差会增大到非常大的程度，远远大于镜

面误差。对于存在折射面的光学系统，这方面的研究还较少。Winston 分析了垂直入射折射面的误差传递，给出了误差传递计算式，但是，他所讨论的系统比较简单，不存在一个方向上的坡度误差向另一个与其垂直方向上光线误差的传递(Winston, Miñano et al. 2005)。

本文根据几何光学原理,从理论上进一步分析了任意一种折射表面的光学误差传递到折射光线误差的关系,给出了光学误差传递到折射光线的误差计算通用表达式,应用该一般表达式分析得到了 8 种常见太阳能聚光系统的计算式。

1、一般原理

测量光学误差时，通常测量的是镜面法线或镜面倾角在两个相互垂直面上投影的角度，计算他们与理想镜面系统的偏差。对于镜面上任一点 P 来说，假设其法线在两个面上投影与坐标轴 x 轴和 y 轴的夹角分别是 $w_x$, 和 $w_y$，镜面误差为 $dw_x$, 和 $dw_y$，将坐标系的 z 轴与理想镜面法向量重合后，法向量误差的计算式为：

$$dn = (w_x, w_y, 0); \tag{3}$$

根据折定理我们有：

$$\vec{t} = \frac{n_i}{n_t}[(-i \cdot n + \sqrt{(i \cdot n)^2 + (n_t/n_i)^2 - 1})\vec{n} + \vec{i}] \tag{4}$$

假设折射线在 xz 面上投影与 x 轴夹角为 θx，在 yz 面上投影与 y 轴夹角为 θy，则

$$\tan\theta x = t_x/t_z \tag{5}$$

两边取导，我们得到：

$$d\theta_x = \frac{t_z dt_x - t_x dt_z}{t_x^2 + t_z^2} \tag{6}$$

$$d\theta_y = \frac{t_z dt_y - t_y dt_z}{t_y^2 + t_z^2} \tag{7}$$

将（4）和（5）式代入到（6）和（7）式，我们得到

$$d\theta_x = (1 - \frac{i_z}{\sqrt{i_z^2 + n_t^2/n_i^2 - 1}})(\delta_x + \frac{i_x i_y}{i_x^2 + i_z^2 + n_t^2/n_i^2 - 1}\delta_y) \tag{8}$$

$$d\theta_y = (1 - \frac{i_z}{\sqrt{i_z^2 + n_t^2/n_i^2 - 1}})(\delta_y + \frac{i_x i_y}{i_y^2 + i_z^2 + n_t^2/n_i^2 - 1}\delta_x) \tag{9}$$

从推导过程可以看出,（9）和（10）是根据光学误差计算折射光线方向误差的一般表达式,适用于所有折射过程。从两式可以看出,折射光线偏差不仅与镜面光学误差相关,还与入射线方向相关,i 是入射线方向矢量。

需要注意的,在应用（9）和（10）式计算时,我们需要将入射光学矢量换算到所在点的法线为 z 轴的坐标系上,x 轴和 y 轴是由镜面误差测量的 x 和 y 方向确定。

如果镜面在两个方向上误差的分布相互独立,标准偏差分别是 $σ_{slopex}$, $σ_{slopey}$, 则反射光线投影在两个方向上的偏差可按下式计算(Bendt, Rabi et al. 1979)：

$$\sigma_x^2 = (1 - \frac{i_z}{\sqrt{i_z^2 + n_t^2/n_i^2 - 1}})^2(\sigma_{slopex}^2 + (\frac{i_x i_y}{i_x^2 + i_z^2 + n_t^2/n_i^2 - 1}\sigma_{slopey})^2) \tag{10}$$

$$\sigma_y^2 = (1-\frac{i_z}{\sqrt{i_z^2+n_t^2/n_i^2-1}})^2(\sigma_{slopey}^2+(\frac{i_x i_y}{i_y^2+i_z^2+n_t^2/n_i^2-1}\sigma_{slopex})^2) \qquad (11)$$

对于平行于误差测量的 x 或 y 方向入射光线，可以简化为：

$$\sigma_x = (1-\frac{i_z}{\sqrt{i_z^2+n_t^2/n_i^2-1}})\sigma_{slopex} = (1-\frac{\tan(\theta_t)}{\tan(\theta_i)})\sigma_{slopex}$$

$$\sigma_y = (1-\frac{i_z}{\sqrt{i_z^2+n_t^2/n_i^2-1}})\sigma_{slopey} = (1-\frac{\tan(\theta_t)}{\tan(\theta_i)})\sigma_{slopey} \qquad (12)$$

这里 $\theta_i$ 和 $\theta_t$ 分别是光线入射角和理论折射角。Winston 等得到的计算折射光线误差计算式(Winston, Miñano et al. 2005)，与我们得到的式 12 是完全相同的。

对于垂直入射的光学系统，ix≈0；iy≈0;iz≈1；两式可以简化：

$$\sigma_x = (1-\frac{n_i}{n_t})\sigma_{slopex}$$

$$\sigma_y = (1-\frac{n_i}{n_t})\sigma_{slopey} \qquad (13)$$

实际折射光学系统存在两个面，需要两次使用计算式分别计算，在按照高斯分布误差累积原理，计算总的折射光线误差。在垂直入射上下两个面（如透明平板）情况下，折射光线总误差为：

$$\sigma_x^2 = (1-\frac{n_t}{n_i})^2\sigma_{slopexdown}^2 + (1-\frac{n_i}{n_t})^2\sigma_{slopexup}^2 \qquad (14)$$

$$\sigma_y^2 = (1-\frac{n_t}{n_i})^2\sigma_{slopeydown}^2 + (1-\frac{n_i}{n_t})^2\sigma_{slopeyup}^2 \qquad (15)$$

我们可以看到，折射材料的折射率影响折射光线误差，折射光线误差随折射率增加而增加。折射率 nt/ni 从 1.5 增加到 3 时，假设上下镜面误差相等，相同镜面误差下，折射光线误差将增加一倍以上。

对于聚光光学系统来说，情况要复杂得多。下面我们分析几种最简单情况。为了简化情况，我们假设各个面在两个方向上的表面误差都相等，定义误差传递系数为反射光线误差与镜面 x 方向或 y 方向坡度误差之比＝σ/σ$_{slope}$。

2、菲涅耳透镜

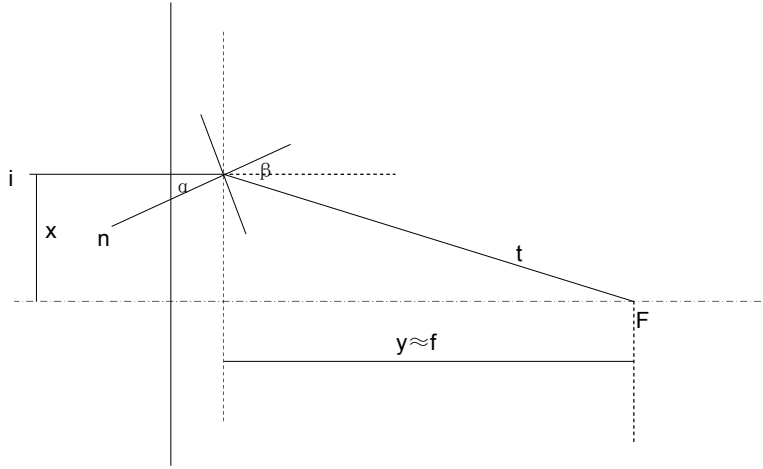

图1 点聚焦菲涅耳透镜误差传递计算示意图

如图是上表面为平面,下表面是斜面的菲涅耳透镜(2003),一束光线 i 经下表面折射到达焦点 F,根据几何光学定理:

$$n_t \sin\alpha = n_a \sin\beta \tag{16}$$

这里 $n_t$ 是折射材料的折射率,$n_a$ 是空气折射率。根据几何位置关系:

$$\tan(\beta-\alpha) = x/y \approx x/f \tag{17}$$

这里 α 是入射角,β 是折射角,x 是折射点距离光轴距离,f 是焦距。令 $n_0 = n_t/n_a$;根据式(16)和(17),我们得到

$$\cos\alpha = \frac{n_0 - \cos(a\tan(x/f))}{\sqrt{(n_0 - \cos(a\tan(x/f)))^2 + (\sin(a\tan(x/f)))^2}} \tag{18}$$

则 i =(sinα,0,cosα),将它们代入到式 10 和 11 中,我们得到:

$$\sigma_x^2 = (1 - \frac{\cos\alpha}{\sqrt{\cos^2\alpha + 1/n_0^2 - 1}})^2 \sigma_{slopexdown}^2 + (1 - \frac{1}{n_0})^2 \sigma_{slopexup}^2 \tag{19}$$

$$\sigma_y^2 = (1 - \frac{\cos\alpha}{\sqrt{\cos^2\alpha + 1/n_0^2 - 1}})^2 \sigma_{slopeydown}^2 + (1 - \frac{1}{n_0})^2 \sigma_{slopeyup}^2 \tag{20}$$

从 20 和 19 式可以看出,对点聚焦菲涅耳透镜,镜面不同位置上的光学误差传递到聚焦光线上的误差是不同的,主要与透镜材料折射率和折射点到主轴距离与焦距之比相关。图 2 是假设透镜上下两个表面误差相同,根据 19 式计算得到的,不同折射率下,聚焦光线误差传递系数(聚焦光线标准偏差与表面误差的标准偏差之比)与折射点到主轴距离比焦距之比(x/f), 从图中可以看出,误差传递系数随 x/f 增大而增大,开始增加缓慢,到 x/f 增大到比较大的时候,就急剧增加,对常用的折射系数为 1.5 左右的玻璃来说,x/f 增大到 0.9 时,聚焦光线误差就增大到非常大的程度了。

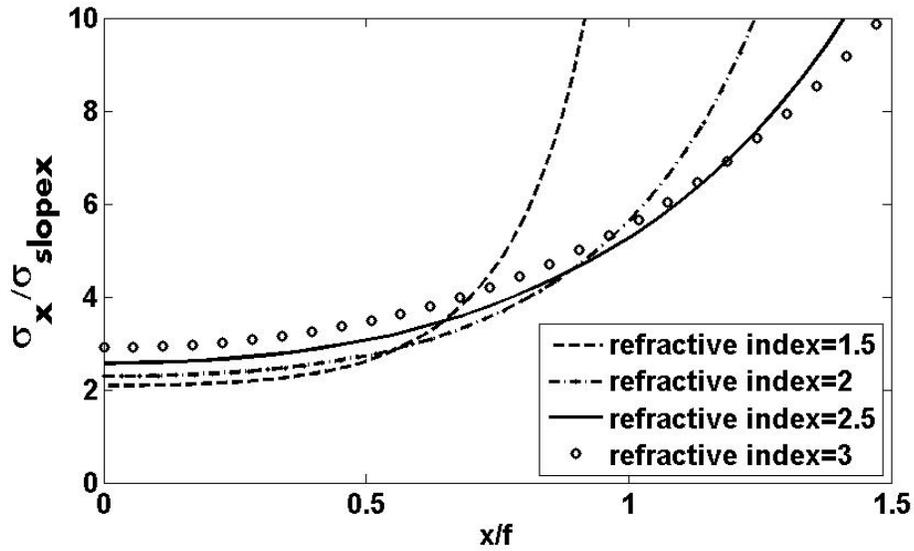

图 2 不同折射率下，误差传递系数与菲涅耳透镜上透射点到主轴距离的关系

3、点聚焦背板反射镜聚光系统

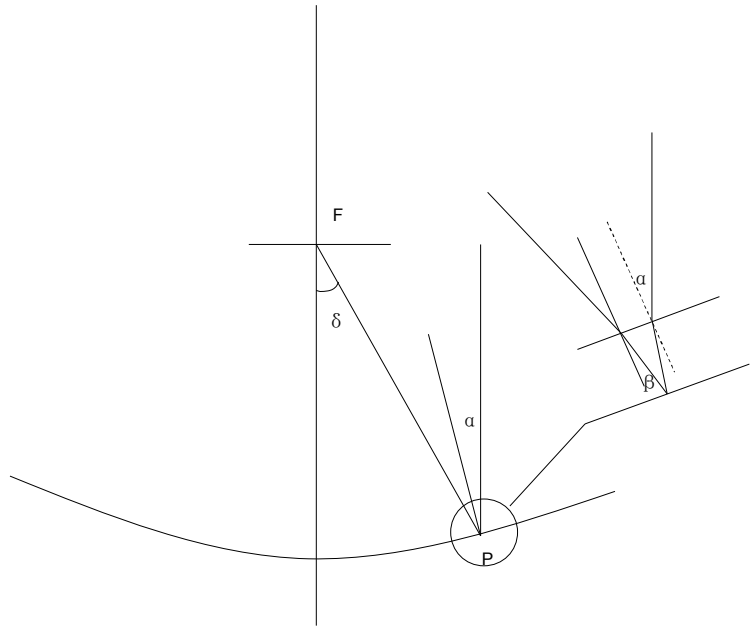

figure 3 旋转抛物面碟式太阳能聚光系统反射光线误差计算示意图

目前最常用的反射镜是玻璃背面镀银的玻璃镜，入射光线先经玻璃折射，再被玻璃背面反射层反射，再经玻璃到空气的折射后聚焦(Kennedy and Terwilliger 2005)。如图，考虑一束平行光轴的光线在旋转抛物面反射镜上 P 点被反射到焦点上，忽略镜面厚度影响，我们有：

$\tan(\alpha)=\tan(\delta/2)=x/(2f)$; (21)

$n_a \sin\alpha = n_t \sin\beta$; (22)

这里 x 是 P 点到反射镜主轴距离，f 是反射镜焦距，对于光线从空气中入射到透明材料上，其入射角为 α，我们有

$$i = (\sin\alpha, 0, \cos\alpha), \quad (23)$$

对于光线从透明材料入射返回大气中，其入射角为 β，我们有：

$$i = (\sin\beta, 0, \cos\beta), \quad (24)$$

代入到式 10 和 11，我们就得到：

$$\sigma_x^2 = (1 - \frac{\cos\alpha}{\sqrt{\cos^2\alpha + 1/n_0^2 - 1}})^2 \sigma_{slopexup}^2 + [(1 - \frac{\cos\beta}{\sqrt{\cos^2\beta + 1/n_0^2 - 1}})^2 + 4]\sigma_{slopexdown}^2 \quad (25)$$

$$\sigma_y^2 = (1 - \frac{\cos\alpha}{\sqrt{\cos^2\alpha + 1/n_0^2 - 1}})^2 \sigma_{slopeyup}^2 + [(1 - \frac{\cos\beta}{\sqrt{\cos^2\beta + 1/n_0^2 - 1}})^2 + 4]\sigma_{slopeydown}^2 \quad (26)$$

式中最后一项是下反射面的误差贡献。从上式可以看出，对点聚焦玻璃反射镜来说，镜面不同位置上的光学误差传递到聚焦光线上的误差是不同的，主要与玻璃折射率和反射点的边缘角相关。图 2 是假设镜面上下两个表面误差相同，根据 25 式计算得到的，不同玻璃折射率下，聚焦光线误差传递系数（聚焦光线标准偏差与表面误差的标准偏差之比）与反射点边缘角的关系，从图中可以看出，误差传递系数随边缘角增大而增大，开始增加缓慢，到边缘角增大到接近 180 度时，就急剧增加，对常用的折射系数为 1.5 左右的玻璃来说，边缘角增大到 2rad 时，增加量都比较小，而反射镜使用折射率较大的透明材料时，误差传递系数随边缘角的增加就比较明显了。与单纯只有反射面的反射镜相比(Huang 2011)，误差传递系数增加幅度与边缘角有关，边缘角较小时，增加较少，而边缘角接近 pi 时，增加很大，其原因在于两次折射带来了误差。

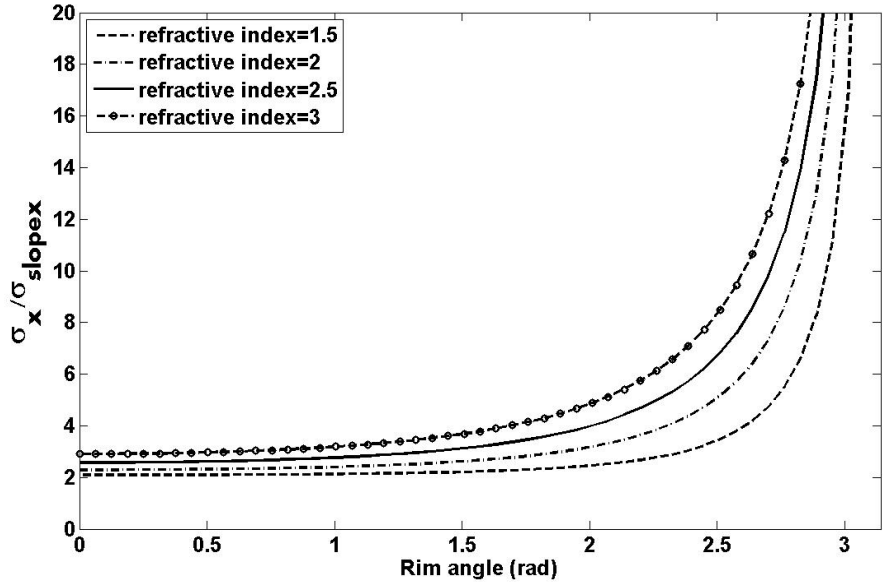

图 4 旋转抛物面玻璃银镜误差传递系数与玻璃折射率和反射点边缘角关系

4、背反射镜抛物面线聚焦系统

槽式系统(Pitz-Paal, Dersch et al. 2007; Grena 2010)是聚光太阳能的主要技术之一。

这里考虑的抛物面槽式系统使用的反射镜是玻璃银镜，在入射点坐标系下，对于入射角为 λ，边缘角为 φ 的入射线矢量为：

$$i =（\cos\lambda\sin(\phi/2),\sin\lambda,\cos\lambda\cos(\phi/2)） \tag{27}$$

该束光线从空气中进入反射镜材料折射带来的误差为：

$$\sigma_{x1}^2 = (1 - \frac{\cos\lambda\cos(\varphi/2)}{\sqrt{\cos^2\lambda\cos^2(\varphi/2) + n_0^2 - 1}})^2(\sigma_{slopexup}^2 + (\frac{\sin\lambda\cos\lambda\sin(\varphi/2)}{\cos^2\lambda + n_0^2 - 1}\sigma_{slopeyup})^2) \tag{28}$$

作为反射面和折射面的入射光线矢量为：

$$\vec{t} = \frac{1}{n_0}(\cos\lambda\sin(\varphi/2),\sin\lambda,\sqrt{\cos^2\lambda\cos^2(\varphi/2) + n_0^2 - 1}) \tag{29}$$

经背面反射层反射后增加的误差为：

$$\sigma_{x2}^2 = 4\sigma_{slopexdown}^2 + (\frac{2\sin\lambda\cos\lambda\sin(\varphi/2)}{\cos^2\lambda + n_0^2 - 1})^2\sigma_{slopeydown}^2 \tag{30}$$

通过透明材料折射返回大气，增加的误差为：

$$\sigma_{x3}^2 = (1 - \frac{\sqrt{\cos^2\lambda\cos^2(\varphi/2) + n_0^2 - 1}}{\cos\lambda\cos(\varphi/2)})^2(\sigma_{slopexup}^2 + (\tan\lambda\sin(\varphi/2))^2\sigma_{slopeyup}^2) \tag{31}$$

反射光线总误差为：

$$\sigma_x^2 = \sigma_{x1}^2 + \sigma_{x2}^2 + \sigma_{x3}^2 \tag{32}$$

上述结果表面，线聚焦槽式玻璃反射镜的误差传递系数与入射角，反射点边缘角以及玻璃折射系数相关。图4给出了根据上式计算得到的槽式抛物面玻璃银镜误差传递系数与入射角和反射点边缘角关系，其中玻璃折射率为1.5，假设镜面误差在x和y方向相同。从结果可以看出，传递系数随折射率增大而增大，也随边缘角增大而增大，在边缘角接近180度时，传递系数增加到非常大。与单纯只有反射面的反射镜相比(Huang 2011)，在边缘角小于pi/2及入射角小于pi/3时，使用玻璃反射镜的误差传递系数增长缓慢，折射带来的影响很小。

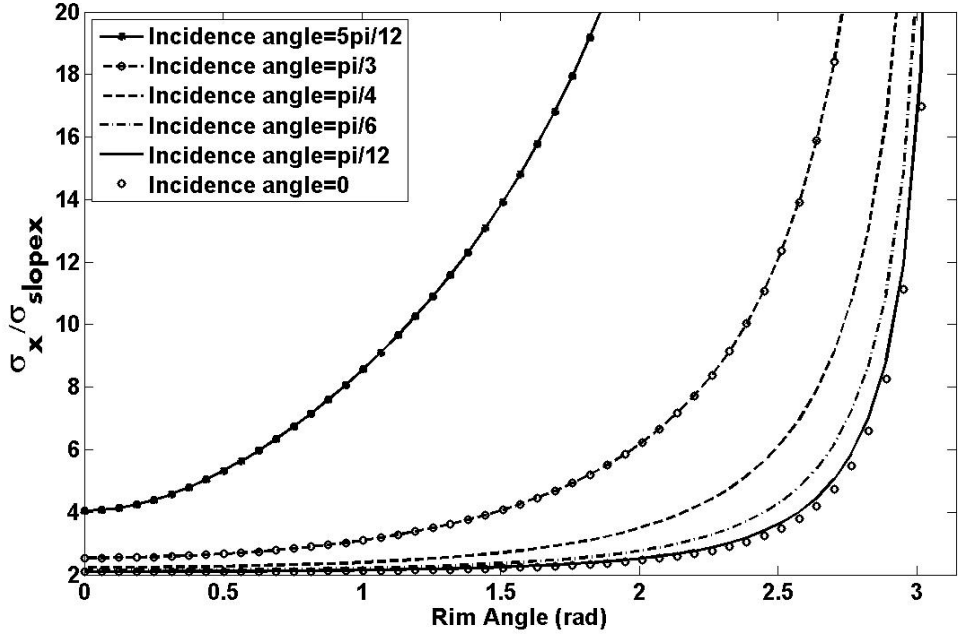

图 5 槽式抛物面玻璃银镜误差传递系数与入射角和反射点边缘角关系，玻璃折射率为 1.5，x 和 y 方向上的表面误差相等

5、定日镜

塔式太阳能系统使用定日镜，将太阳能反射到中心接收器上，是有发展前景的太阳能利用系统(Yao, Wang et al. 2009)。定日镜常使用玻璃反射镜制作。当入射角为 λ，反射点方位角为 β，边缘角为 α，则入射光线矢量为：

i＝（sin λ cosβcosα-cos λ sinα, sin( λ )sinβ, sin λ cosβsinα+cos λ cosα）
  ≈（sin λ cosβ, sin λ sinβ, cos λ）                                              （33）

由于边缘角很小，我们近似认为 α≈0. 将式 33 代入到式 10 和 11，可得到该束光线从空气到玻璃的折射，误差为：

$$\sigma_{x1}^2 = (1-\frac{\cos\lambda}{\sqrt{\cos^2\lambda+n_0^2-1}})^2(\sigma_{slopexup}^2+(\frac{\sin^2\lambda\sin(2\beta)/2}{(\sin\lambda\cos\beta)^2+\cos^2\lambda+n_0^2-1}\sigma_{slopeyup})^2) \quad (34)$$

$$\sigma_{y1}^2 = (1-\frac{\cos\lambda}{\sqrt{\cos^2\lambda+n_0^2-1}})^2(\sigma_{slopeyup}^2+(\frac{\sin^2\lambda\sin(2\beta)/2}{(\sin\lambda\sin\beta)^2+\cos^2\lambda+n_t^2/n_a^2-1}\sigma_{slopexup})^2)$$

(35)

反射光线和折射的入射光线矢量为：

$$\vec{t} = \frac{1}{n_0}(\sin\lambda\cos\beta,\sin\lambda\sin\beta,\sqrt{\cos^2\lambda+n_0^2-1}) \quad (36)$$

经背面反射层反射后增加的误差为：

$$\sigma_{x2}^2 = 4\sigma_{slopexdown}^2 + (\frac{\sin^2\lambda\sin(2\beta)}{\sin^2\lambda\cos^2\beta+\cos^2\lambda+n_0^2-1})^2\sigma_{slopeydown}^2 \quad (37)$$

$$\sigma_{y2}^2 = 4\sigma_{slopeydown}^2 + (\frac{\sin^2\beta\sin(2\lambda)}{\cos^2\lambda + \sin^2\lambda\sin^2\beta + n_0^2 - 1})^2\sigma_{slopexdown}^2 \tag{38}$$

再通过透明材料折射返回大气,增加的误差为:

$$\sigma_{x3}^2 = (1 - \frac{\sqrt{\cos^2\lambda + n_0^2 - 1}}{\cos\lambda})^2(\sigma_{slopexup}^2 + (\frac{0.5\sin^2\lambda\sin(2\beta)}{\sin^2\lambda\cos^2\beta + \cos^2\lambda}\sigma_{slopeyup})^2) \tag{39}$$

$$\sigma_{y3}^2 = (1 - \frac{\sqrt{\cos^2\lambda + n_0^2 - 1}}{\cos\lambda})^2(\sigma_{slopeyup}^2 + (\frac{0.5\sin^2\lambda\sin(2\beta)}{\cos^2\lambda + \sin^2\beta\sin^2\lambda}\sigma_{slopexup})^2) \tag{40}$$

反射光线总误差为:

$$\sigma_x^2 = \sigma_{x1}^2 + \sigma_{x2}^2 + \sigma_{x3}^2 \tag{41}$$

$$\sigma_y^2 = \sigma_{y1}^2 + \sigma_{y2}^2 + \sigma_{y3}^2 \tag{42}$$

上述结果表面,定日镜反射光线在任意一个方向上的误差,不仅与镜面两个方向上的光学误差相关,还与入射角和玻璃折射系数相关;以及反射点在镜面的方位角有关。图 6 和 7 给出了根据计算式计算得到的定日镜在 x 方向和 y 方向误差传递系数与入射角和反射点方位角关系,其中玻璃折射率为 1.5。从结果可以看出,传递系数主要随入射角增大而增大;方位角对传递系数影响,仅在入射角较大时,才会有较大影响。与单纯只有反射面的反射镜相比(Huang 2011),在入射角小于等于 pi/4 时,玻璃反射镜略微增大,其原因在于两次折射带来了额外误差。在入射角较大时,误差传递系数增加幅度较大。

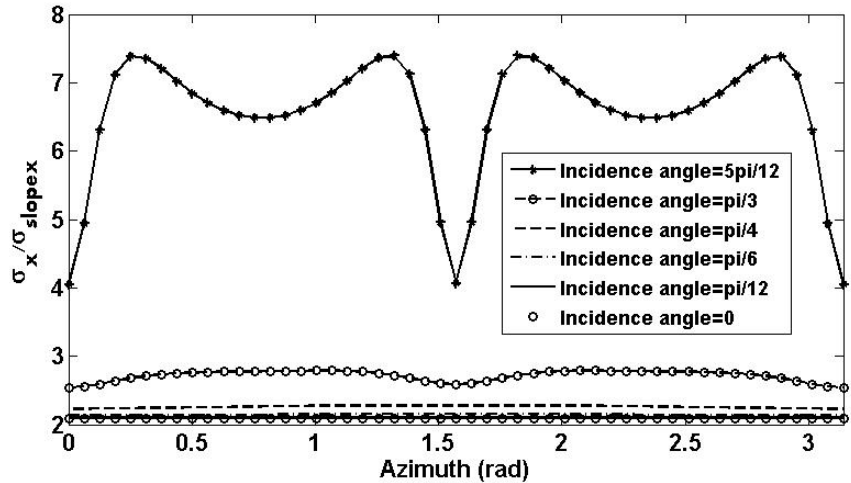

图 6 玻璃银镜做定日镜在 x 方向误差传递系数与入射角和反射点边缘角关系,玻璃折射率为 1.5,各光学面及其在 x 和 y 方向上的坡度误差均相等

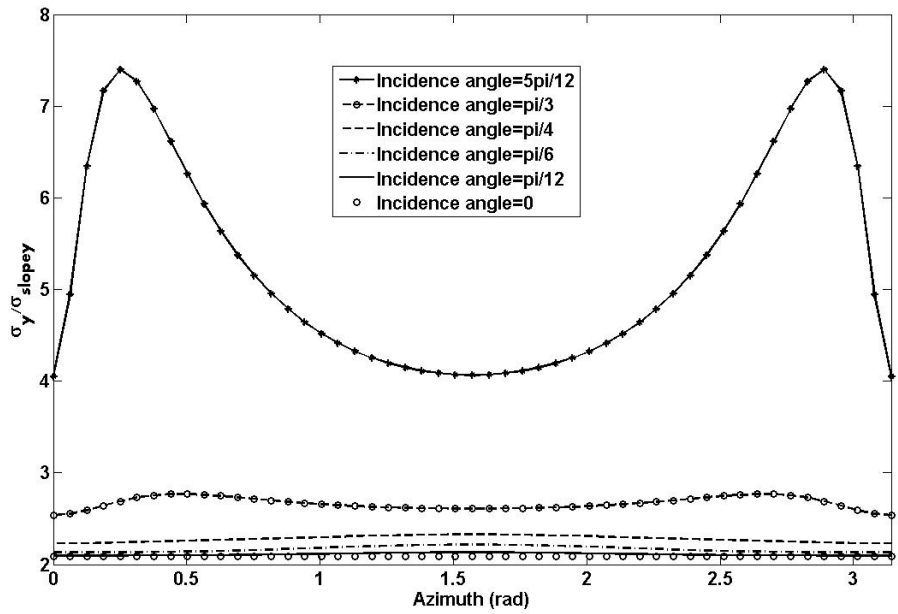

图 7 玻璃银镜做定日镜在 y 方向误差传递系数与入射角和反射点边缘角关系，玻璃折射率为 1.5，各光学面及其在 x 和 y 方向上的坡度误差均相等

6、棱镜全反射聚光 TIR-R concentrator

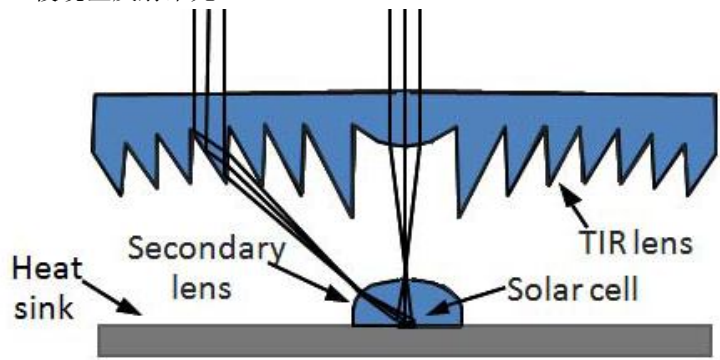

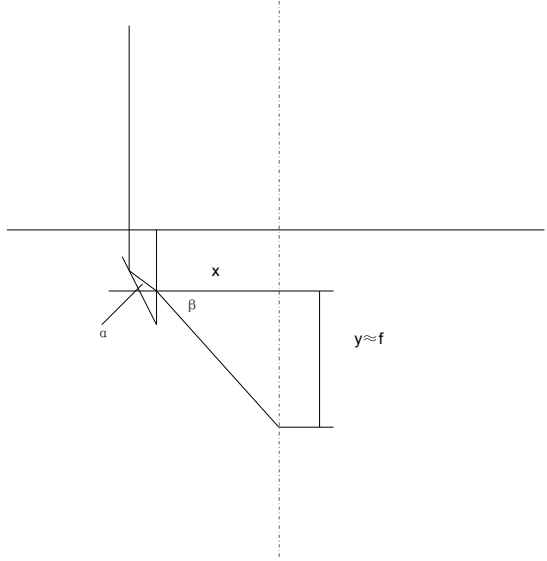

图 8 棱镜全反射菲涅耳聚光镜及其误差传递系数计算

棱镜全反射菲涅耳聚光镜与普通菲涅耳透镜结构类似，不同之处在于利用棱镜全反射作用聚焦(Hernandez, Benitez et al. 2003)。如图 8。在全反射模式下，入射光线经两次折射和 1 次反射，棱镜出射面与系统主轴平行，所以出射面上的入射角可按下式计算：

$n_t \sin\alpha = n_a \sin\beta$ （43）

$\tan\beta = y/x \approx f/x$ （44）

根据 43 和 44 式，我们得到：

$$\cos\alpha = \sqrt{1 - \frac{f^2}{x^2 + f^2}\frac{n_a^2}{n_t^2}}$$ （45）

总误差计算与菲涅耳透镜类似，为：

$$\sigma_x^2 = (1-\frac{1}{n_0})^2 \sigma_{slopex1}^2 + 4\sigma_{slopex2}^2 + (1 - \frac{\cos\alpha}{\sqrt{\cos^2\alpha + 1/n_0^2 - 1}})^2 \sigma_{slopex3}^2$$

$$= (1-\frac{1}{n_0})^2 \sigma_{slopex1}^2 + 4\sigma_{slopex2}^2 + (1 - \sqrt{n_0^2 - (n_0^2-1)\frac{f^2}{x^2}})^2 \sigma_{slopex3}^2$$ （46）

$$\sigma_y^2 = (1-\frac{1}{n_0})^2 \sigma_{slopey1}^2 + 4\sigma_{slopey2}^2 + (1 - \frac{\cos\alpha}{\sqrt{\cos^2\alpha + 1/n_0^2 - 1}})^2 \sigma_{slopey3}^2$$

$$= (1-\frac{1}{n_0})^2 \sigma_{slopey1}^2 + 4\sigma_{slopey2}^2 + (1 - \sqrt{n_0^2 - (n_0^2-1)\frac{f^2}{x^2}})^2 \sigma_{slopey3}^2$$ （47）

误差传递系数，在 x 和 y 方向上是相等的，主要与镜面聚光点到主轴距离与焦距比值相关，在比值较小时，误差传递系数很大，这与该系统主轴周围一般使用普通透镜是一致的；随着比值增大而逐渐减小，两个折射面的误差传递趋向于光学垂直于平板情况。

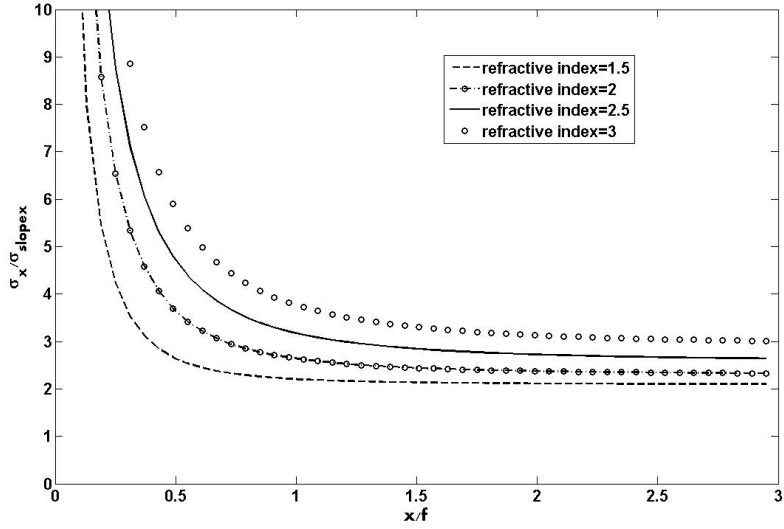

图 9 棱镜全反射菲涅耳聚光透镜误差传递系数与反射点位置关系，玻璃折射率为 1.5，各光学面及其在 x 和 y 方向上的坡度误差均相等

7、线聚焦菲涅耳系统

通常线聚焦菲涅耳系统常使用圆柱形玻璃背面镀反射膜的圆柱面镜(Mills and Morrison 2000)。假设入射光线在系统长度方向上入射角为 λy，在另一个方向上的入射角为 λx，则对镜面上任一点 P 来说，以镜面对称轴为 z 轴的坐标系下，入射光线矢量为：

$$i = (\tan(\lambda x), \tan(\lambda y), 1)/\sqrt{\tan^2(\lambda x)+\tan^2(\lambda y)+1} \quad (48)$$

以 P 点法线为 z 轴的坐标系是上述坐标系绕 y 轴旋转 $-\alpha$，所以，在这个坐标系下，入射光线矢量为：

$$i = (\tan(\lambda x)\cos\alpha - \sin\alpha, \tan(\lambda y), \tan(\lambda x)\sin\alpha+\cos\alpha)/\sqrt{\tan^2(\lambda x)+\tan^2(\lambda y)+1} \quad (49)$$

对于线聚焦塔式系统来说，通常焦距很大，而镜面开口宽度较小，可以近似认为：$\sin\alpha=0$，$\cos\alpha=1$。所以我们得到入射光线矢量：

$$i = (\tan(\lambda x), \tan(\lambda y), 1)/\sqrt{\tan^2(\lambda x)+\tan^2(\lambda y)+1} \quad (50)$$

**将其代入计算式 10 可得到：**

$$\sigma_{x1}^2 = (1- \frac{1}{\sqrt{(n_0^2-1)(\tan^2\lambda_x+\tan^2\lambda_y)+n_0^2}})^2(\sigma_x^2+(\frac{\tan\lambda_x \tan\lambda_y}{\tan^2\lambda_x+(n_0^2-1)(\tan^2\lambda_x+\tan^2\lambda_y)+n_0^2})^2\sigma_y^2)$$

(51)

光线在反射面上入射角和从玻璃到空气的入射角均为：

$$i_2 = (i_x, i_y, \sqrt{i_z^2+n_0^2-1})/n_0 \quad (52)$$

根据反射面误差传递公式和上式，反射面误差带来的聚焦光线误差影响：

$$\sigma_{x2}^2 = 4\sigma_{slopex}^2 + (\frac{\tan\lambda_x \tan\lambda_y}{\tan^2\lambda_x+(n_0^2-1)(\tan^2\lambda_x+\tan^2\lambda_y)+n_0^2})^2\sigma_{slopey}^2 \quad (53)$$

将式代入到 10 式，可得到光线在反射镜上表面出射折射带来的影响：

$$\sigma_{x3}^2 = (1-\sqrt{(n_0^2-1)(\tan^2\lambda_x + \tan^2\lambda_y) + n_0^2})^2(\sigma_x^2 + (\sin\lambda_x\cos\lambda_x\tan\lambda_y)^2\sigma_y^2) \qquad (54)$$

反射光线总误差为：

$$\sigma_x^2 = \sigma_{x1}^2 + \sigma_{x2}^2 + \sigma_{x3}^2 \qquad (55)$$

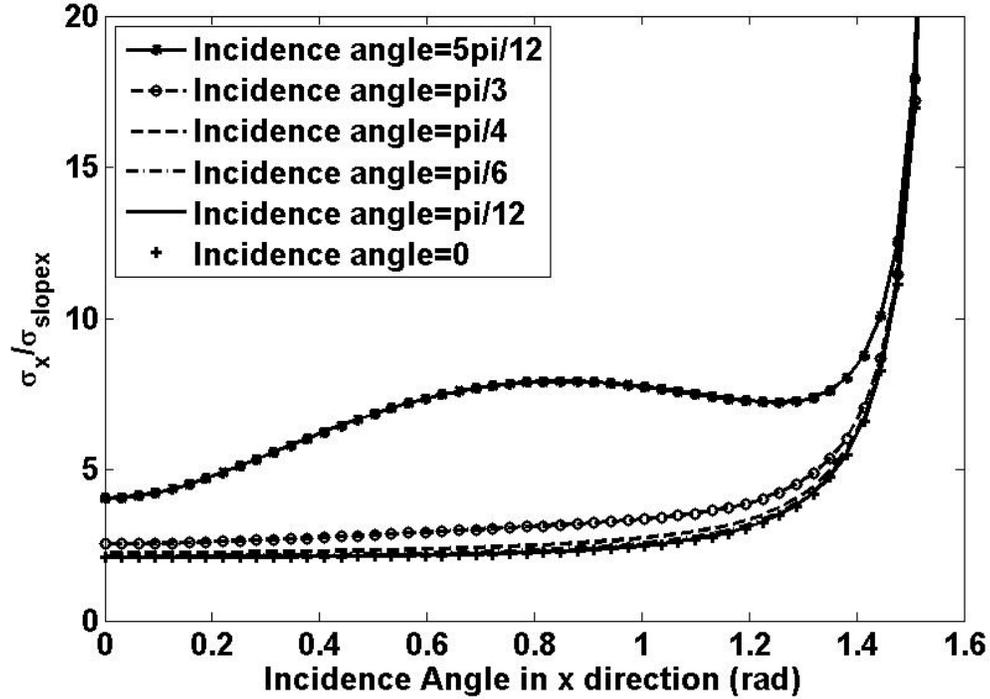

图 10 线聚焦塔式系统光学误差对反射线 x 方向误差的传递，假设光学误差在 x 和 y 方向相同，图中入射角是 y 方向入射角

假设光学误差在 x 和 y 方向相同情况下，根据上式得到的计算结果如图 7。我们可以看到，在相同 y 方向入射角下，随着 x 方向入射角的增加，基本上反射光线误差会增大；y 方向入射角增大，反射光线误差也会增加，但在 y 方向入射角大于 pi/4 时，随入射角增大，误差传递系数明显增大；

通常线聚焦塔式系统安装在南北方向跟踪轴，在高纬度地区，冬天中午太阳的高度角较小，例如，在北纬 40 度地区，冬至日中午太阳高度角仅 26.55º，计算得到的误差传递系数至少增加到 2.70； x 方向入射角为 pi/4，增大到 3.5，这说明，对高纬度地区冬季来说，反射光线误差增大，性能明显下降，但其增长幅度低于采用反射膜的反射镜(Huang 2011)。

8、卡塞格林式反射聚焦

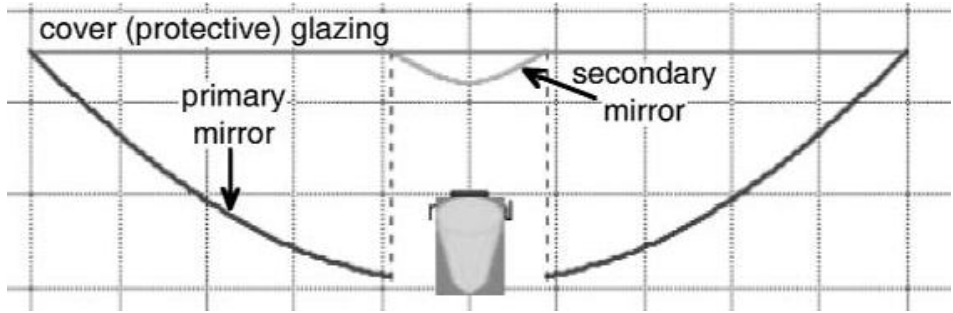

卡塞格林式反射聚焦是聚光太阳能电池的主要结构形式之一(Luque and Andreev 2007)。如图所示，系统有主副两个反射镜和光伏电池组成。主镜是抛物面反射镜，次镜是双曲线反射镜，假设两个反射镜均使用背板式反射镜，主镜误差传递到反射光线误差的计算同抛物面反射镜(式25和26)，副镜误差传递到反射光线上的误差计算，参见下图，主镜焦点和副镜上焦点重合，接收器放置在副镜下焦点上，假设主镜和副镜顶点间距离为d，副镜两焦点间距离为2c，来自主镜边缘角为 δ 光线入射到副镜上的入射角为θ，我们设双曲线方程为：

$$x^2/a^2 - y^2/b^2 = 1; \tag{56}$$

则 $a=c-f+d$; $b^2=c^2-a^2$，入射光线方程为：

$$y = \tan(180-\delta)(x-c) \tag{57}$$

根据两式可以得到入射线在副镜上反射点坐标（x0，y0），则入射角 θ

$$\theta = \text{atan}(2x_0 y_0/c^2 - x_0^2 - y_0^2) \tag{58}$$

对折射角 φ，我们有

$$n_a \sin\theta = n_t \sin\varphi \tag{59}$$

所以副镜光线误差带来的反射光线误差为：

$$\sigma_x^2 = [(1 - \frac{\cos\varphi}{\sqrt{\cos^2\varphi + 1/n_0^2 - 1}})^2 + (1 - \frac{\cos\theta}{\sqrt{\cos^2\theta + n_0^2 - 1}})^2]\sigma_{slopexup}^2 + 4\sigma_{slopexdown}^2 \tag{60}$$

$$\sigma_y^2 = [(1 - \frac{\cos\varphi}{\sqrt{\cos^2\varphi + 1/n_0^2 - 1}})^2 + (1 - \frac{\cos\theta}{\sqrt{\cos^2\theta + n_0^2 - 1}})^2]\sigma_{slopeyup}^2 + 4\sigma_{slopeydown}^2 \tag{61}$$

此外，我们还需要加上玻璃盖板的误差（式14和15）。

9、卡塞格林透镜

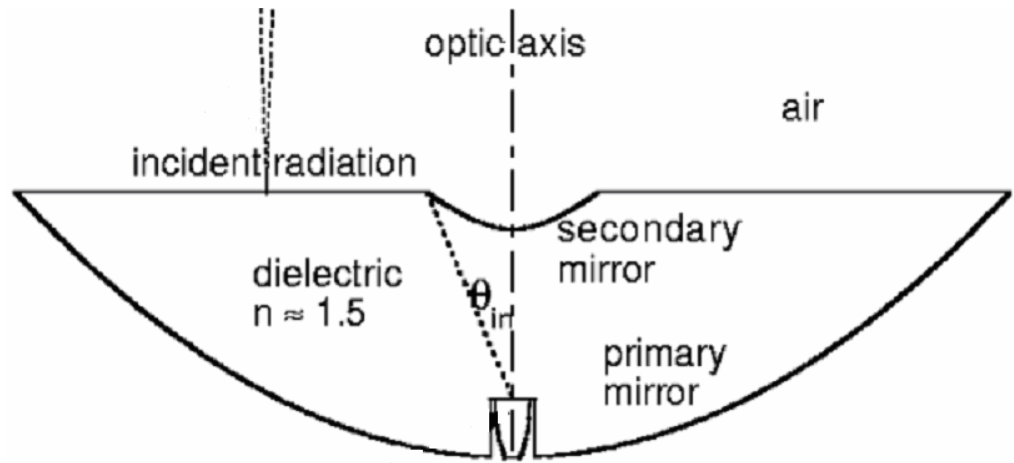

卡塞格林透镜是上下两个面为反射镜的透镜(Luque and Andreev 2007)，光线通过上下两个反射面经过两次折射和 2 次反射，从底部出射到达接受器。两次折射中，从空气中入射的折射是垂直入射的，从介质中折射到空气中，也可近似看成是垂直入射，或按几何光学原理，根据第一次反射点的边缘角计算入射角 $\theta_{in}$，具体可参见卡塞格林式反射聚焦，所以，总误差为：

$$\sigma_x{}^2 = (1-1/n_0)^2 \sigma_{x1}{}^2 + 4\sigma_{x2}{}^2 + 4\sigma_{x3}{}^2 + (1-\frac{\cos\theta_{in}}{\sqrt{\cos^2\theta_{in}+1/n_0^2-1}})^2 \sigma_{x4}{}^2 \quad （62）$$

$$\sigma_y{}^2 = (1-1/n_0)\sigma_{y1}{}^2 + 4\sigma_{y2}{}^2 + 4\sigma_{y3}{}^2 + (1-\frac{\cos\theta_{in}}{\sqrt{\cos^2\theta_{in}+1/n_0^2-1}})^2 \sigma_{y4}{}^2 \quad （63）$$

10、 讨论和结论

本文根据几何光学理论,推导了任意镜面上光学误差传递给折射光线产生的误差计算通用公式，公式表明，传递到折射光线上的误差，不仅与镜面光线误差相关，而且与入射光线方向和透镜折射率相关。折射光线误差传递系数随折射率增加而增加，也随入射角增大而增加。

我们应用该计算式分析了 8 种聚光太阳能系统情况，包括点聚焦菲涅耳透镜，点聚焦抛物面玻璃反射镜，线聚焦抛物面玻璃反射镜，使用玻璃银镜的塔式系统定日镜，棱镜全反射菲涅耳聚光镜，使用玻璃反射镜的线性菲涅耳系统，使用玻璃银镜的卡塞格林式反射聚焦系统，及卡塞格林透镜。结果显示，即使镜面光学误差相同，由于误差传递系数与入射角和入射位置等相关，各种聚光镜上不同位置上的反射光线误差是不同的，点聚焦菲涅耳透镜，点聚焦抛物面玻璃反射镜， 使用玻璃反射镜的抛物槽，线聚焦菲涅耳玻璃反射镜的误差传递系数随反射点距离轴的距离的增加而增加，而棱镜全反射菲涅耳聚光镜则相反，随距离增加而减小。使用玻璃反射镜的定日镜主要与入射角和方位角相关。对槽式反射镜和定日镜，还存在 y 方向光学误差传递到光线 x 方向现象，放大了聚焦光线误差。

在扫描光学系统(Yu and Tan 2003)中，广泛存在大入射角情况，从而可能存在误差传递带来的误差放大问题。在系统设计时，考虑本文提出的误差传递规律，是非常必要的。

参考文献：